\documentclass[prl,twocolumn,aps,amsmath,amssymb,showpacs,10pt]{revtex4-1}

\usepackage{amsbsy}
\usepackage{graphicx}
\usepackage{bm}
\usepackage{graphics}
\usepackage{latexsym}

\usepackage[active]{srcltx}

\newcommand{\sect}[1]{\setcounter{equation}{0}\section{#1}}

\def\spazio#1{\vrule height#1em width0em depth#1em}

\def\spazio#1{\vrule height#1em width0em depth#1em}

\begin{document}

\begin{abstract}
{This paper aims at proving the fundamental role of a relativistic formulation for quarkonia models.
 We present a completely covariant description of a two-quark system interacting by the Cornell potential with a 
Breit term describing  the hyperfine splitting. Using an appropriate procedure to calculate the Breit correction, we
find heavy meson masses in excellent agreement with experimental data. Moreover, also when
applied to light quarks and even taking average values of the running coupling constant, we prove that covariance
properties and hyperfine splitting are sufficient to explain the light mesons spectrum and to give a very good
agreement with the data. 
 }
\end{abstract}

\bigskip
\bigskip

\title{\bf Unified covariant treatment of hyperfine splitting for heavy and light mesons.}
%from relativistic two fermion wave equation.}

\author{R. Giachetti}
%\email[Email: ]{giachetti@fi.infn.it}
\affiliation{Dipartimento di Fisica, Universit\`a di Firenze,
Italy} \affiliation{Istituto Nazionale di Fisica Nucleare, Sezione
di Firenze} 
\author{E. Sorace}
%\email[Email: ]{sorace@fi.infn.it}
\affiliation{Istituto Nazionale di Fisica Nucleare, Sezione di
Firenze}

\maketitle

\bigskip

%
%%%%%%%%%%%%%%%%%%%%%%%%%%%%%%%%%%%%%%%%%%%%%%%%%%%
%
%                        TEXT
%
%%%%%%%%%%%%%%%%%%%%%%%%%%%%%%%%%%%%%%%%%%%%%%%%%%%
%

%%%%%%%%%%%%%%%%%%%%%%%%%%%%%%%%%%%%%%%%%%%%%%%%%%%

\sect{Introduction} \label{Sec_introduction}

%%%%%%%%%%%%%%%%%%%%%%%%%%%%%%%%%%%%%%%%%%%%%%%%%%%

Potential models of interacting quark systems have a long history and are still a very lively subject of
investigation: this is witnessed by the large  number  of research papers and reviews that keep being published
\cite{Rich}, which we refer to for bibliography and  exhaustive details on the subject.
Since the first papers that gave a rather complete overall picture of the subject \cite{LSGI}, the 
starting point is often a Schr\"odinger equation with a potential having a Coulomb behavior at the
origin and confining at infinity; the relativistic corrections together with the
spin-orbit and the spin-spin contributions are taken into account by adding terms which are
treated perturbatively. Attempts have also been made to overcome the limitations of a potential model due
to asymptotic freedom at short distances and to light quark creation: a description of these effects has been
tried by means of screened potentials softening the Coulomb interaction at the origin and by letting the confining
term saturate at infinity. 
The spin dependent interactions are then modeled by the Breit-Fermi potential with a $\delta$-function
centered at the origin, which in many cases yields difficulties in explaining the hyperfine splittings of the spectra.
Although this approximation may be good for heavy mesons, a smearing of the $\delta$-function has been proposed 
 to get a better description of the small distance behavior: recent results \cite{RR-BBD}, however, show that this
point has not been settled. 

A major point of discussion has always been the relevance of relativistic properties of the systems,
not only in the obvious case of light mesons, but also for heavy mesons. A truly
covariant formulation going beyond the ``relativized'' treatment has often been invoked and approaches in such
direction have been actually worked out  \cite{RK-CS,Bra,SeCe,HC}.  Many of them are
connected with field theory along the lines of the Bethe-Salpeter equation and the spectra of the resulting equations
are not of straightforward computation. Few models deal with a consistent relativistic description. 
In \cite{Bra} a full spinor treatment is presented. The confinement is essentially obtained by a cutoff of the
wave function at a fixed interparticle separation, the Breit interaction is differently treated for light
and heavy mesons and an \emph{ad hoc} contact interaction is introduced: the approach is interesting but
not fully covariant. A covariant formulation is given in \cite{SeCe}; however, since the main subject of
investigation are the Regge trajectories, the assumed potential is just linear in the radial variable. The  
papers in \cite{HC} study a well formulated relativistic model with a two-body Dirac equation derived from
constraint dynamics. The interaction is first introduced by a relativistic extension of the Adler-Piran
potential and then improved  by the addition of a time-like confining vector potential, 
yielding very good results. 

We present here a canonical description of quarkonium, focusing on the complete covariance of the formulation and on
the fermionic nature of the elementary constituents. 
The formulation originates from a wave equation for two relativistic fermions with arbitrary masses   
obtained from two Dirac operators coupled by the interaction \cite{GS2}. We refer to those papers for the proofs of the
full covariance, of the Schr\"odinger and the one-particle Dirac limits, as well as of the cyclicity of the relative
time that avoids the difficulties of relative energy excitations. 
We observe that our construction has different assumptions from \cite{HC}, so that the final equations and the results
also are somewhat different.
In \cite{GS2}  the hyperfine splitting
of Positronium was calculated, finding an agreement better than up to the fourth power of the fine structure constant
with the results obtained by QED semi-classical expansions. 
In the present context we will use the simplest Cornell potential with a Breit term for the spin-spin interaction. 
Our purpose is to show that the full relativistic description and a proper perturbation treatment of the  Breit term,
avoiding the evaluation of a delta function at the origin, are already sufficient to give results in excellent
agreement with the experimental data both for heavy and light mesons, contrary to some diffused ideas. 
Further improvements of the potential are an
important issue which should be developed at a  more phenomenological level of the investigation. For instance in
our calculations we have used average values of the running coupling constant (rcc) for the different families of
mesons, verifying \emph{ex post} that the ratios of the assumed values are in agreement with those obtained from the 
well known $\alpha_S$ curve \cite{pdg}: a fine tuning of the rcc, modeled according to the $\alpha_S$ curve,
should produce much better results.

%
%
%%%%%%%%%%%%%%%%%%%%%%%%%%%%%%%%%%%%%%%%%%%%%%%%%%%
%%%%%%%%%%%%%%%%%%%%%%%%%%%%%%%%%%%%%%%%%%%%%%%%%%%
%%%%%%%%%%%%%%%%%%%%%%%%%%%%%%%%%%%%%%%%%%%%%%%%%%%
%
%\smallskip
\begin{table}[b]
{{ $~~~$
\begin{tabular}{lcc}
  \hline
  $~~~~~~${\texttt{State}}$\phantom{{}^{{}^{\displaystyle{i}}}}$ &$\texttt{Exp}$ 
&$ \texttt{Num}$
   \\
  \hline
%
%   1s
%
  %
$({\texttt{1}}^{\texttt{1}}{\texttt{s}}_{\texttt{0}})~0^+(0^{-+})~
\phantom{{}^{{}^{\displaystyle{i}}}}\!\! \eta_b$
&$\phantom{XX}$\texttt{9390.90$\pm$2.8}$\phantom{XX}$  
&\texttt{\phantom{1}9390.39}
 \\ 
%\hline
  $({\texttt{1}}^{\texttt{3}}{\texttt{s}}_{\texttt{1}})~0^-(1^{--})~
\phantom{{}^{{}^{\displaystyle{i}}}}\!\!  \varUpsilon$
&\texttt{9460.30$\pm$.25}  
&\texttt{\phantom{1}9466.10}
 \\ 
%\hline
%
%   1p
%
  $({\texttt{1}}^{\texttt{3}}{\texttt{p}}_{\texttt{0}})~0^+(0^{++})~
\phantom{{}^{{}^{\displaystyle{i}}}}\!\!  \chi_{b0}$
&\texttt{9859.44$\pm$.73}  
&\texttt{\phantom{1}9857.41}
 \\ %\hline
  $({\texttt{1}}^{\texttt{3}}{\texttt{p}}_{\texttt{1}})~0^+(1^{++})~
\phantom{{}^{{}^{\displaystyle{i}}}}\!\!  \chi_{b1}$
&\texttt{9892.78$\pm$.57}  
&\texttt{\phantom{1}9886.70}$\spazio{0.5 } $
 \\ %\hline
  $({\texttt{1}}^{\texttt{1}}{\texttt{p}}_{\texttt{1}})~0^-(1^{+-})~
\phantom{{}^{{}^{\displaystyle{i}}}}\!\!  h_b$
&\texttt{9898.60$\pm$1.4}  
&\texttt{\phantom{1}9895.35}
 \\ %\hline
  $({\texttt{1}}^{\texttt{3}}{\texttt{p}}_{\texttt{2}})~0^+(2^{++})~
\phantom{{}^{{}^{\displaystyle{i}}}}\!\!  \chi_{b2}$
&\texttt{9912.21$\pm$.57}  
&\texttt{\phantom{1}9908.14}$\spazio{0.5 } $
 \\ 
%\hline
%
%   2s
%
 %\hline
  $({\texttt{2}}^{\texttt{3}}{\texttt{s}}_{\texttt{1}})~0^-(1^{--})~
\phantom{{}^{{}^{\displaystyle{i}}}} \!\! \varUpsilon$
&\texttt{10023.26$\pm$.0003}  
&\texttt{10009.04}
 \\ 
%\hline
%
%   2d
%
  $({\texttt{1}}^{\texttt{3}}{\texttt{d}}_{\texttt{2}})~0^-(2^{--})~
\phantom{{}^{{}^{\displaystyle{i}}}}\!\!  \varUpsilon_2$
&\texttt{10163.70$\pm$1.4}  
&\texttt{10152.69}
 \\ 
 %\hline
%
%   2p
%
  $({\texttt{2}}^{\texttt{3}}{\texttt{p}}_{\texttt{0}})~0^+(0^{++})~
\phantom{{}^{{}^{\displaystyle{i}}}}\!\!  \chi_{b0}$
&\texttt{10232.50$\pm$.0009}  
&\texttt{10232.36}
 \\ 
  %\hline
  $({\texttt{2}}^{\texttt{3}}{\texttt{p}}_{\texttt{1}})~0^+(1^{++})~
\phantom{{}^{{}^{\displaystyle{i}}}}\!\!  \chi_{b1}$
&\texttt{10255.46$\pm$.0005}  
&\texttt{10256.58}
 \\ 
   %\hline
  $({\texttt{2}}^{\texttt{3}}{\texttt{p}}_{\texttt{2}})~0^+(2^{++})~
\phantom{{}^{{}^{\displaystyle{i}}}}\!\!  \chi_{b2}$
&\texttt{10268.65$\pm$.0007}  
&\texttt{10274.26}
 \\
%\hline
%
%   3s
  %\hline
  $({\texttt{3}}^{\texttt{3}}{\texttt{s}}_{\texttt{1}})~0^-(1^{--})~
\phantom{{}^{{}^{\displaystyle{i}}}}\!\!  \varUpsilon$
&\texttt{10355.20$\pm$.0005}  
&\texttt{10364.52}
 $\spazio{0.5}$\\ 
 %\hline
  $({\texttt{3}}^{\texttt{3}}{\texttt{p}}_{\texttt{0}})~0^+(0^{++})~
\phantom{{}^{{}^{\displaystyle{i}}}} \!\! \chi_{b0}$
&\texttt{}  
&\texttt{10534.86}\\
 
 %\hline
  $({\texttt{3}}^{\texttt{3}}{\texttt{p}}_{\texttt{1}})~0^+(1^{++})~
\phantom{{}^{{}^{\displaystyle{i}}}} \!\! \chi_{b1}$
&$<$\texttt{10530$\pm$.014}$>_{{}_J}$  
&\texttt{10556.59}\\
  %\hline
 %\hline
  $({\texttt{3}}^{\texttt{3}}{\texttt{p}}_{\texttt{2}})~0^+(2^{++})~
\phantom{{}^{{}^{\displaystyle{i}}}} \!\! \chi_{b2}$
&\texttt{}  
&\texttt{10572.44}\\ 
  %\hline
%  $~$\texttt{-----}
   %
%&\texttt{-----}  
   %
%&\texttt{-----}
% \\ 
  $({\texttt{4}}^{\texttt{3}}{\texttt{s}}_{\texttt{1}})~0^-(1^{--})~
\phantom{{}^{{}^{\displaystyle{i}}}} \!\! \varUpsilon$
&\texttt{10579.40$\pm$.0012}  
&\texttt{10655.34}\\ 
  %\hline
  $({\texttt{5}}^{\texttt{3}}{\texttt{s}}_{\texttt{1}})~0^-(1^{--})~
\phantom{{}^{{}^{\displaystyle{i}}}}\!\!  \varUpsilon$
&\texttt{10876$\pm$11}  
&\texttt{10910.35}
$\spazio{0.5}$\\ 
\hline
  \end{tabular}
}}
\label{Table_b-bbar}
\caption{The \texttt{b}\texttt{\={b}} levels in MeV. First column: term symbol,
$I^G(JPC)$ numbers , particle name. $\sigma$=$\texttt{1.111}\,$GeV/fm,
$\alpha$=\texttt{0.3272}, $m_b$=$\texttt{4725.5}\,$MeV. Experimental data from \cite{pdg}.
}
\end{table}
% 
%
%%%%%%%%%%%%%%%%%%%%%%%%%%%%%%%%%%%%%%%%%%%%%%%%%%%
%%%%%%%%%%%%%%%%%%%%%%%%%%%%%%%%%%%%%%%%%%%%%%%%%%%
%%%%%%%%%%%%%%%%%%%%%%%%%%%%%%%%%%%%%%%%%%%%%%%%%%%
%
%

\sect{The two-fermion wave equation with Cornell potential and Breit term} \label{Sec_equation}

The Dirac operators entering the wave equation prescribe the correct form for the interactions
according to their tensorial nature: the Coulomb-like term of the Cornell potential is vectorial and thus
minimally coupled to the energy; the linear term is scalar and therefore coupled to the mass. Indeed
only a scalar growing potential is actually confining, while an unbounded vector interaction is not
\cite{Pl-GSG}. We refer to \cite{GS2} for the derivation of the radial system of the model. 
We call $r_a$, $q_a$ the Wigner vectors of spin one given by the spatial parts of relative
coordinates and momenta boosted to the frame with vanishing total spatial momentum
 and we put $r=(r_ar_a)^{1/2}$ (sum over repeated indexes).  We denote by $\gamma_{{(i)}}$ the gamma
matrices acting in the spinor space of the $i$-th fermion of mass $m_{(i)}$, $M=m_{(1)}+m_{(2)}$
and $\rho=\left|m_{(1)}-m_{(2)}\right |/M$.
The vector and scalar couplings produce the terms
$E+b/r$,  $\frac12(M+\sigma r)$ and the final wave equation reads
\begin{eqnarray}
&{}&\!\!\!\Bigl[\,\Bigl(
{\gamma}^0_{(1)}{\gamma_{(1)}}_{a}-
{\gamma}^0_{(2)}{\gamma_{(2)}}_{a}\Bigr)q_a+\frac12\Bigl({\gamma}^0_{(1)}\!
+\! {\gamma}^0_{(2)}\Bigr)\Bigl(M+\sigma r\Bigr)+\spazio{1.2}\cr
&{}&\phantom{ii}\frac12\Bigl({\gamma}^0_{(1)}\!
-\! {\gamma}^0_{(2)}\Bigr)M\rho -\Bigl(E+\frac b r\Bigr)+ V_B(r)
\,\Bigr]\,\Psi(\vec{r})=0.
%\nonumber
\label{BreitHam}
\end{eqnarray}
where  
\begin{eqnarray}
V_B(r)=\frac b{2r}\,
{\gamma}^0_{(1)}{\gamma_{(1)}}_{a}{\gamma}^0_{(2)}{\gamma_{(2)}}_{b}
\Bigl(\delta_{ab}\!+\!\frac{r_ar_b}{r^2}\Bigr) 
\label{BreitTerm}
\end{eqnarray}
is the Breit term generating the hyperfine splitting. As in \cite{GS2} the first perturbation order
of this term is evaluated by
substituting $V_B(r)$ with $\varepsilon V_B(r)$ in (\ref{BreitHam}) and taking the first derivative of the 
eigenvalues with respect to $\varepsilon$  in $\varepsilon\!=\!0$ from the numerical solutions of the differential
equations. This could also be seen as an application of the spectral correspondence to the Feynman-Hellman
theorem.

The radial system is obtained by diagonalizing angular momentum and parity. As in \cite{GS2} it is formed by four
algebraic plus four first order differential equations for each parity.
Using the algebraic relations and defining the dimensionless variables
$\,\Omega\,,\,w\,,s\,$ by 
\begin{eqnarray}
\!\!\!\sigma=\frac {M^2}4\,\Omega^{\frac32},~~ E=\frac M2\,(2+\Omega w),~~ r=\frac 2M\,\Omega^{-\frac12}\,s,
\label{Dimensionless_Variables}
%\nonumber 
\end{eqnarray}
the radial system for (\ref{BreitHam}), replacing $V_B(r)$ by $\varepsilon V_B(r)$, is
\begin{eqnarray}
\!\left(\!\! {\begin{array}{c}
                                          u'_1(s)\\
                                          u'_2(s)\\
                                          u'_3(s)\\
                                          u'_4(s)
                                         \end{array}}\!\!\right)\!\!+\!\! 
\left( \!\!{\begin{array}{cccc} 
0 &\!\!\!\phantom{-}A_0(s) &\!\!\!\,\, -B_0(s) &\!\!\! \phantom{-}0\\
A_\varepsilon(s) &\!\!\! \phantom{-}{ {1}/{s}}  &\!\!\! \phantom{-}0 &\!\!\!   B_\varepsilon(s) \\ 
{C}_\varepsilon(s) &\!\!\!\phantom{-}0 &\!\!\! \phantom{-} { {2}/{s}}  &\!\!\! \phantom{-}A_\varepsilon(s)\\ 
0 &\!\!\!  {D}_\varepsilon(s) &\!\!\!\phantom{-}A_0(s) &\!\!\! \phantom{-}{ {1}/{s}}
\end{array}}
\!\! \right)\!\!
\left(\!\! {\begin{array}{c}
                                          u_1(s)\\
                                          u_2(s)\\
                                          u_3(s)\\
                                          u_4(s)
                                         \end{array}}\!\!\right)\!= 0.
\label{System}
\nonumber
\end{eqnarray}
Here $A_0\!=\!A_\varepsilon|_{\varepsilon=0}$,
$B_0\!=\!B_\varepsilon|_{\varepsilon=0}\,$ and $u'(s)\!=\!du(s)/ds$.
Letting $\,J^2\!=\!j(j+1)$, the even parity coefficients  are:
\begin{eqnarray}
&{}&A_\varepsilon(s)= \frac{2\sqrt{J^2}\,\rho}{\sqrt{\Omega}\,(sh(s)-2\varepsilon b)}\,,\spazio{1.0}\cr
&{}& B_\varepsilon(s)= \frac{(h^2(s)/2-2\rho^2/\Omega)\,s^2-2\varepsilon^2b^2}{s^2h(s)-2\varepsilon bs}
\,,\spazio{1.0}\cr 
&{}&C_\varepsilon(s)=\frac{h(s)}2 +
\frac{2\varepsilon b}s+\frac{2J^2}{2\varepsilon b s-s^2h(s)}+\frac{2s\,k^2(s)}{4\varepsilon b-s\,h(s)}
\,,\spazio{1.0}\cr
&{}&D_\varepsilon(s)=\frac{2J^2}{s^2h(s)}-\frac{4b^2\varepsilon^2-s^2h^2(s)+4s^2k^2(s)}{4\varepsilon b
s-2s^2h(s)}
\label{CFeven}
%\nonumber
\end{eqnarray}
with 
$h(s)=(2+\Omega w)/\sqrt{\Omega}+ b/s,\,\,  k(s)=(2+\Omega s)/(2\sqrt{\Omega}).$ 
%
%%%%%%%%%%%%%%%%%%%%%%%%%%%%%%%%%%%%%%%%%%%%%%%%%%%
%
\begin{table}[b]
{{ $~~~$
\begin{tabular}{lcc}
  \hline
  $~~~~~~${\texttt{State}}$\phantom{{}^{{}^{\displaystyle{i}}}}$ &$\texttt{Exp}$ 
&$ \texttt{Num}$
   \\
  \hline
%
%   1s
%
$({\texttt{1}}^{\texttt{1}}{\texttt{s}}_{\texttt{0}})~0^+(0^{-+})~
\phantom{{}^{{}^{\displaystyle{i}}}}\!\! \eta_c$
&$\phantom{XX}$\texttt{2978.40$\pm$1.2}$\phantom{XX}$  
&\texttt{2978.26}
 \\ 
  $({\texttt{1}}^{\texttt{3}}{\texttt{s}}_{\texttt{1}})~0^-(1^{--})~
\phantom{{}^{{}^{\displaystyle{i}}}}\!\! J/\psi$
&\texttt{3096.916$\pm$.011}  
&\texttt{3097.91}
 \\ 
%
%   1p
%
  $({\texttt{1}}^{\texttt{3}}{\texttt{p}}_{\texttt{0}})~0^+(0^{++})~
\phantom{{}^{{}^{\displaystyle{i}}}}\!\! \chi_{c0}$
&\texttt{3414.75$\pm$.31}  
&\texttt{3423.88}
 \\ 
  $({\texttt{1}}^{\texttt{3}}{\texttt{p}}_{\texttt{1}})~0^+(1^{++})~
\phantom{{}^{{}^{\displaystyle{i}}}}\!\! \chi_{c1}$
&\texttt{3510.66$\pm$.07}  
&\texttt{3502.83}$\spazio{0.5 } $
 \\ 
  $({\texttt{1}}^{\texttt{1}}{\texttt{p}}_{\texttt{1}})~0^-(1^{+-})~
\phantom{{}^{{}^{\displaystyle{i}}}}\!\! h_c$
&\texttt{3525.41$\pm$.16}  
&\texttt{3523.67}
 \\ 
  $({\texttt{1}}^{\texttt{3}}{\texttt{p}}_{\texttt{2}})~0^+(2^{++})~
\phantom{{}^{{}^{\displaystyle{i}}}}\!\! \chi_{c2}$
&\texttt{3556.20$\pm$.09}  
&\texttt{3555.84}$\spazio{0.5 } $
 \\ 
%
%   2s
%
  $({\texttt{2}}^{\texttt{1}}{\texttt{s}}_{\texttt{0}})~0^+(0^{-+})~
\phantom{{}^{{}^{\displaystyle{i}}}}\!\! \eta_c$
&\texttt{3637$\pm$4}  
&\texttt{3619.64}
 \\ 
   $({\texttt{2}}^{\texttt{3}}{\texttt{s}}_{\texttt{1}})~0^-(1^{--})~
\phantom{{}^{{}^{\displaystyle{i}}}} \!\!\psi$
&\texttt{3686.09$\pm$.04}  
&\texttt{3692.91}
 \\ 
%
%   2d
%
  $({\texttt{1}}^{\texttt{3}}{\texttt{d}}_{\texttt{1}})~0^-(1^{--})~
\phantom{{}^{{}^{\displaystyle{i}}}}\!\! \psi$
&\texttt{3772.92$\pm$.35}  
&\texttt{3808.48}
 \\ 
%
%   2p
%
  $\phantom{({\texttt{?}}^{\texttt{?}}{\texttt{?}}_{\texttt{?}})}~0^+(?^{?+})~
\phantom{{}^{{}^{\displaystyle{i}}}}${\small{\texttt{X(3872)}}} 
&\texttt{3871.57$\pm$.25}  
&\texttt{}
 \\
  %hline
  $({\texttt{2}}^{\texttt{3}}{\texttt{p}}_{\texttt{1}})~0^+(1^{++})~
\phantom{{}^{{}^{\displaystyle{i}}}}\!\! \chi_{c1}$ 
&\texttt{-}  
&\texttt{3961.21}
 \\ 
  $\phantom{({\texttt{?}}^{\texttt{?}}{\texttt{?}}_{\texttt{?}})}~0^+(?^{?+})~
\phantom{{}^{{}^{\displaystyle{i}}}}${\small{\texttt{X(3915)}}} 
&\texttt{3917.4$\pm$2.7}  
&\texttt{}
 \\
%
%   2p
%
  $({\texttt{2}}^{\texttt{3}}{\texttt{p}}_{\texttt{2}})~0^+(2^{++})~
\phantom{{}^{{}^{\displaystyle{i}}}}\!\! \chi_{c2}$
&\texttt{3927$\pm$2.6}  
&\texttt{4003.93}
 \\
  $\phantom{({\texttt{?}}^{\texttt{?}}{\texttt{?}}_{\texttt{?}})}~?^+(?^{??})~\,
\phantom{{}^{{}^{\displaystyle{i}}}}${\small{\texttt{X(3940)}}}
&\texttt{3942$\pm$13}  
&\texttt{}\\
%
%   3s
%
  $({\texttt{3}}^{\texttt{1}}{\texttt{s}}_{\texttt{0}})~0^+(0^{-+})~
\phantom{{}^{{}^{\displaystyle{i}}}} \!\!\eta_c$
&\texttt{-}  
&\texttt{4064.21}
 \\ 
  $({\texttt{3}}^{\texttt{3}}{\texttt{s}}_{\texttt{1}})~0^-(1^{--})~
\phantom{{}^{{}^{\displaystyle{i}}}}\!\! \psi$
&\texttt{4039$\pm$1}  
&\texttt{4122.95}\\
  $({\texttt{2}}^{\texttt{3}}{\texttt{d}}_{\texttt{1}})~0^-(1^{--})~
\phantom{{}^{{}^{\displaystyle{i}}}}\!\! \psi$
&\texttt{4153$\pm$3}  
&\texttt{4200.51}\\
  $({\texttt{4}}^{\texttt{3}}{\texttt{s}}_{\texttt{1}})~0^-(1^{--})~
\phantom{{}^{{}^{\displaystyle{i}}}}\!\! \psi$
&\texttt{4421$\pm$4}  
&\texttt{4479.22}
 $\spazio{0.5}$\\    
\hline
  \end{tabular}
}}
\label{Table_c-cbar}
\caption{The \texttt{c}\texttt{\={c}} levels in MeV. $\sigma$=$\texttt{1.111}\,$GeV/fm, 
$\alpha$=\texttt{0.435}, $m_c$=$\texttt{1394.5}\,$MeV. Experimental data from \cite{pdg}.
}
\end{table}
% 
%%%%%%%%%%%%%%%%%%%%%%%%%%%%%%%%%%%%%%%%%%%%%%%%%%%
%
The coefficients for the odd parity system are: 
\begin{eqnarray}
&{}&A_\varepsilon(s)= \frac{2\,\sqrt{J^2}\,k(s)}{2\varepsilon b-s\,h(s)},
\spazio{1.0}\cr
&{}&B_\varepsilon(s)=\frac{4\varepsilon^2b^2-s^2h^2(s)+4s^2k^2(s)}{4\varepsilon b
s-2s^2h(s)} \,,\spazio{1.0}\cr
&{}&C_\varepsilon(s)=\frac{h(s)}2\!+\!\frac{2J^2}{2\varepsilon bs-s^2h(s)}\!
+\!\frac{2\varepsilon b}{s}\!+\!\frac{2s\rho^2}{\Omega(4\varepsilon b-sh(s))} \,,\spazio{1.0}\cr
&{}&D_\varepsilon(s)=-\frac{h(s)}2+\frac{2J^2}{s^2h(s)}-\frac{\varepsilon b}{s}
+\frac{2\rho^2s}{\Omega\,(sh(s)-2b s)}
%\nonumber
\label{CFodd}
\end{eqnarray}

A word about the numerical method we have used is in order. The origin and infinity are the only singular points of the
 boundary value problem and no further singularities arise from the  matrix of the coefficients.. The solution was
obtained by a double shooting method, the spectral condition being the vanishing of the $4\!\times\!4$ determinant of
the matching conditions at a crossing point \cite{GS2}. 
Pad\'e techniques have been used to improve the accuracy of the approximate solutions at  zero and infinity. The
integration precision has always been kept very high and tested
against the stability of the spectral values.

%\smallskip
\begin{table}[t]
{{ $~~~$
\begin{tabular}{lcc}
  \hline
  $~~~~~~${\texttt{State}}$\phantom{{}^{{}^{\displaystyle{i}}}}$ &$\texttt{Exp}$ 
&$ \texttt{Num}$
   \\
  \hline
%
%   1s
%
  %
$({\texttt{1}}^{\texttt{1}}{\texttt{s}}_{\texttt{0}})~0^+(0^{-+})~
\phantom{{}^{{}^{\displaystyle{i}}}}\!\! $
&$\phantom{XX}$\texttt{-}$\phantom{XX}$  
&\texttt{\phantom{1}818.12}
 \\ 
  $({\texttt{1}}^{\texttt{3}}{\texttt{s}}_{\texttt{1}})~0^-(1^{--})~
\phantom{{}^{{}^{\displaystyle{i}}}}\!\! \phi$
&\texttt{1019.455$\pm$.020}  
&\texttt{1019.44}
 \\ 
%
%   1p
%
  $({\texttt{1}}^{\texttt{3}}{\texttt{p}}_{\texttt{1}})~0^+(1^{++})~
\phantom{{}^{{}^{\displaystyle{i}}}}\!\!${\small{\texttt{f${}_{\texttt{1}}$(1420)}}}
&\texttt{ 1426.4$\pm$.9}  
&\texttt{1412.84}$\spazio{0.5 } $
 \\ 
  $({\texttt{1}}^{\texttt{3}}{\texttt{p}}_{\texttt{2}})~0^+(2^{++})~
\phantom{{}^{{}^{\displaystyle{i}}}}\!\!${\small{\texttt{f${}'${}${}_{\texttt{2}}$(1525)}}}
&\texttt{1525$\pm$5}  
&\texttt{1525.60}$\spazio{0.5 } $
 \\ 
%
%   2s
%
  $({\texttt{2}}^{\texttt{3}}{\texttt{s}}_{\texttt{1}})~0^-(1^{--})~
\phantom{{}^{{}^{\displaystyle{i}}}}\!\! \phi$
&\texttt{1680$\pm$20}  
&\texttt{1698.41}
 \\ 
%
%   2d
%
  $({\texttt{1}}^{\texttt{3}}{\texttt{d}}_{\texttt{1}})~0^-(1^{--})~
\phantom{{}^{{}^{\displaystyle{i}}}}\!\!${\small{\texttt{X(1750)}}}
&\texttt{1753.5$\pm$3.8}  
&\texttt{1776.53}
 \\ 
$({\texttt{1}}^{\texttt{3}}{\texttt{d}}_{\texttt{3}})~0^-(3^{--})~
\phantom{{}^{{}^{\displaystyle{i}}}}\!\! ${$\phi_3${\texttt{(1850)}}}
&\texttt{1854$\pm$7}  
&\texttt{1880.85}
 \\
%
%   2p
%
  $({\texttt{2}}^{\texttt{3}}{\texttt{p}}_{\texttt{2}})~0^+(2^{++})~
\phantom{{}^{{}^{\displaystyle{i}}}}\!\! ${\small{\texttt{f${}_{\texttt{2}}$(2010)}}}
&\texttt{2011$\pm$70}  
&\texttt{2073.15}
 \\
%
%   3s
%
  $({\texttt{3}}^{\texttt{3}}{\texttt{s}}_{\texttt{1}})~0^-(1^{--})~
\phantom{{}^{{}^{\displaystyle{i}}}}\!\! \phi$
&\texttt{2175$\pm$15}  
&\texttt{$\,$2217.57}
$\spazio{0.5}$\\ 
\hline
  \end{tabular}
}}
\label{Table_s-sbar}
\caption{The \texttt{s}\texttt{\={s}} levels in MeV. $\sigma$=1.34 GeV/fm, 
$\alpha$=\texttt{0.6075}, $m_s$=$\texttt{134.27}\,$MeV. Experimental data from \cite{pdg}. 
}
\end{table}
%
%%%%%%%%%%%%%%%%%%%%%%%%%%%%%%%%%%%%%%%%%%%%%%%%%%%
%

% 
%%%%%%%%%%%%%%%%%%%%%%%%%%%%%%%%%%%%%%%%%%%%%%%%%%%
%
%\smallskip
\begin{table}[b]
{{ $~~~$
\begin{tabular}{lcc}
  \hline
  $~~~~~~${\texttt{State}}$\phantom{{}^{{}^{\displaystyle{i}}}}$ &$\texttt{Exp}$ 
&$ \texttt{Num}$
   \\
 \hline
$({\texttt{1}}^{\texttt{1}}{\texttt{s}}_{\texttt{0}})~0(0^{-})~
\phantom{{}^{{}^{\displaystyle{i}}}}\!\! ${\small{\texttt{B${}_{\texttt{c}}^{\texttt{$\pm$}}$}}}
&\texttt{6277$\pm$.006}  
&\texttt{6277}
 \\ 
\hline
%
%   1s
%
  %
$({\texttt{1}}^{\texttt{1}}{\texttt{s}}_{\texttt{0}})~0(0^{-})~
\phantom{{}^{{}^{\displaystyle{i}}}}\!\! ${\small{\texttt{B${}_{\texttt{s}}^{\texttt{0}}$}}}
&\texttt{5366.77$\pm$.24}  
&\texttt{5387.41}
 \\ 
  $({\texttt{1}}^{\texttt{3}}{\texttt{s}}_{\texttt{1}})~0(1^{-})~
\phantom{{}^{{}^{\displaystyle{i}}}}\!\! ${\small{\texttt{B${}_{\texttt{s}}^{\texttt{*}}$}}}
&\texttt{5415.4$\pm$2.1}  
&\texttt{5434.34}
 \\ 
%
%   1p
%
  $({\texttt{1}}^{\texttt{3}}{\texttt{p}}_{\texttt{1}})~0(1^{+})~
\phantom{{}^{{}^{\displaystyle{i}}}}\!\!${\small{\texttt{B${}_{\texttt{s1}}$(5830)${}^{\texttt{0}}$}}}
&\texttt{ 5829.4$\pm$.7}  
&\texttt{5817.80}$\spazio{0.5 } $
 \\ 
  $({\texttt{1}}^{\texttt{3}}{\texttt{p}}_{\texttt{2}})~0(2^{+})~
\phantom{{}^{{}^{\displaystyle{i}}}}\!\!${\small{\texttt{B${}_{\texttt{s2}}$(5840)${}^{\texttt{0}}$}}}
&\texttt{5839.7$\pm$.6}  
&\texttt{5829.33}$\spazio{0.5 } $\ 
\smallskip
 \\
%--------------------------------\phantom{-} &---------------\phantom{-} &\phantom{-}------------\phantom{-}
 \hline
$({\texttt{1}}^{\texttt{1}}{\texttt{s}}_{\texttt{0}})~0(0^{-})~
\phantom{{}^{{}^{\displaystyle{i}}}}\!\! ${\small{\texttt{D${}_{\texttt{s}}^\pm$}}}
&\texttt{1968.49$\pm$.32}  
&\texttt{1961.24}
 \\ 
$({\texttt{1}}^{\texttt{3}}{\texttt{s}}_{\texttt{1}})~0(1^{-})~
\phantom{{}^{{}^{\displaystyle{i}}}}\!\! ${\small{\texttt{D${}_{\texttt{s}}^{*\pm}$}}}
&\texttt{2112.3$\pm$.50}  
&\texttt{2101.78}
 \\ 
  $({\texttt{1}}^{\texttt{3}}{\texttt{p}}_{\texttt{0}})~0(0^{+})~
\phantom{{}^{{}^{\displaystyle{i}}}}\!\!${\small{\texttt{D${}_{\texttt{s0}}$(2317)${}^\pm$}}}
&\texttt{2317.8$\pm$.6}  
&\texttt{2339.94}
 \\ 
%
%   2d
%
  $({\texttt{1}}^{\texttt{3}}{\texttt{p}}_{\texttt{1}})~0(1^{+})~
\phantom{{}^{{}^{\displaystyle{i}}}}\!\!${\small{\texttt{D${}_{\texttt{s1}}$(2460)${}^\pm$}}}
&\texttt{2459.6$\pm$.6}  
&\texttt{2466.15}
 \\ 
  $({\texttt{1}}^{\texttt{1}}{\texttt{p}}_{\texttt{1}})~0(1^{+})~
\phantom{{}^{{}^{\displaystyle{i}}}}\!\!${\small{\texttt{D${}_{\texttt{s1}}$(2536)${}^\pm$}}}
&\texttt{2535.12$\pm$.13}  
&\texttt{2535.82}
 \\ 
  $({\texttt{1}}^{\texttt{3}}{\texttt{p}}_{\texttt{2}})~0(2^{+})~
\phantom{{}^{{}^{\displaystyle{i}}}}\!\!${\small{\texttt{D${}_{\texttt{s2}}^{*}$(2573)}}}
&\texttt{2571.9$\pm$.8}  
&\texttt{2574.92}

$\spazio{0.5}$\\ 
\hline
  \end{tabular}
}}
\label{Table_Bc-Bs-Cs}
\caption{The \texttt{Bc}, \texttt{Bs} and \texttt{Ds} levels in MeV. $\sigma$=\texttt{1.111},
\texttt{1.111}, \texttt{1.227} GeV/fm and 
$\alpha$=\texttt{0.3591}, \texttt{0.3975}, \texttt{0.5348} respectively. 
}
\end{table}
%
%%%%%%%%%%%%%%%%%%%%%%%%%%%%%%%%%%%%%%%%%%%%%%%%%%%
%

\sect{Discussion of the numerical results } \label{Sec_bb_cc}
As stated in the Introduction, in order to have a test as good as possible of the relevance of the relativistic dynamics
in quarkonium models, we have aimed at choosing the least number of fit parameters.  
Flavor independence could be expected for heavy quarks. In fact, doing separate fits for $b\bar{b}$, $b\bar{s}$ and
$c\bar{c}$ we
find that the string tensions turn out to be the same within the computation precision. The same values of $\sigma$ and
of the masses are taken for the
unique measured $Bc$ state. We introduce $\alpha=(3/4)\,b$, where $b$ is the parameter of the Cornell potential
appearing in (\ref{CFeven},\ref{CFodd}). We assume a constant $\alpha$ determined by a separate fit for each family of
mesons. However the ratios 
$\alpha_{b\bar{b}}/\alpha_{c\bar{c}}$ = 0.752, 
$\alpha_{b\bar{b}}/\alpha_{b\bar{c}}$ = 0.911,
$\alpha_{b\bar{c}}/\alpha_{b\bar{s}}$ = 0.903,
$\alpha_{b\bar{c}}/\alpha_{c\bar{s}}$ = 0.672,
$\alpha_{c\bar{c}}/\alpha_{s\bar{s}}$ = 0.716,
$\alpha_{s\bar{s}}/\alpha_{u\bar{d}}$ = 0.926,
numerically found,
are very close to the corresponding ratios 
$\alpha_S(\chi_{b1,1P})/\alpha_S(\chi_{c0,1P})=0.754$,
$\alpha_S(\chi_{b1,1P})/\alpha_S(B_c^\pm)=0.914$,
$\alpha_S(B_c^\pm)/\alpha_S(B_s^{*})=0.955$,
$\alpha_S(B_c^\pm)/\alpha_S(D_c^{*\pm})=0.686$,
$\alpha_S(\chi_{c0,1P})/\alpha_S(f_{1,1P})=0.714$, 
$\alpha_S(f_{1,1P})/\alpha_S(a_{1,1P})/=0.933$ 
for average values
$\Lambda_S=0.221,\,0.296,\,0.349\,$ GeV for $n_f=5,4,3$ \cite{pdg}.

The spectra show common features, generally shared by all potential models: the states group into doublets of $s$ states
and quadruplets of $p$, $d, ...$ states. It clearly appears that the results are in very good agreement  with
experimental data below the thresholds of $B$ and $D$ mesons \cite{pdg} for $b\bar b$ and $c\bar c$ respectively. Above
the thresholds the calculated energies of the levels are larger than the experimental ones and a softened potential
could make a sensible difference in reproducing the data of higher levels. The regularity of the
pattern is however maintained. 
From Table II, for instance, as the resonance $X(3782)$ has the two
possible assignments $J^{PC}=1^{++}$ and $2^{-+}$ \cite{pdg}, the model could indicate a $\chi_{c1}$ classification. 
Nothing can be suggested for $X(3915)$ and $X(3940)$, having no accepted quantum numbers.
The situation is simpler in Table I, where there are no unclassified physical states. We point out the good estimate of
the recently discovered $\chi_b(3P)$ resonance \cite{pdg}, staying just below the $B$ production threshold. On the
contrary, the calculated values for $\varUpsilon({\texttt{4}}^{\texttt{3}}{\texttt{s}}_{\texttt{1}})$ and
$\varUpsilon({\texttt{5}}^{\texttt{3}}{\texttt{s}}_{\texttt{1}})$ exceed the experimental values. 
%
%We present in Table III the calculated absolute values of the Breit corrections for some states of $b\bar b$,
%$c\bar c$ and $s\bar s$: they are responsible for the hyperfine splitting and have to be subtracted from the initial
%pure Cornell potential levels in order to obtain the values given in Tables I, II and IV. As expected, the corrections
%decrease for increasing values of $J$ and become more and more important for decreasing masses of the component quarks.
%
%
%%
%\smallskip
\begin{table}[t]
{{ $~~~$
\begin{tabular}{lcc}
  \hline
  $~~~~~~${\texttt{State}}$\phantom{{}^{{}^{\displaystyle{i}}}}$ &$\texttt{Exp}$ 
&$ \texttt{Num}$
   \\
  \hline
%
%   1s
%
  %
%$({\texttt{1}}^{\texttt{1}}{\texttt{s}}_{\texttt{0}})~0^+(0^{-+})~
%\phantom{{}^{{}^{\displaystyle{i}}}}\!\! \eta_b$
   %
%&$\phantom{XX}$\texttt{9390.90$\pm$2.8}$\phantom{XX}$  
   %
%&\texttt{\phantom{1}9390.39}
% \\ 
%\hline
  $({\texttt{1}}^{\texttt{3}}{\texttt{s}}_{\texttt{1}})~1^+(1^{--})~
\phantom{{}^{{}^{\displaystyle{i}}}}\!\! ${\small{\texttt{$\rho$(770)}}}
&\texttt{775.49$\pm$.39}  
&\texttt{\phantom{1}826.14}
 \\ 
%\hline
%
%   1p
%
  $({\texttt{1}}^{\texttt{3}}{\texttt{p}}_{\texttt{0}})~1^-(0^{++})~
\phantom{{}^{{}^{\displaystyle{i}}}}\!\!${\small{\texttt{a${}_{\texttt{0}}$(980)}}}
&\texttt{980.$\pm$20}  
&\texttt{\phantom{11}970.34 }
 \\ %\hline
  $({\texttt{1}}^{\texttt{3}}{\texttt{p}}_{\texttt{1}})~1^-(1^{++})~
\phantom{{}^{{}^{\displaystyle{i}}}}\!\!  ${\small{\texttt{a${}_{\texttt{1}}$(1260)}}}
&\texttt{1230.$\pm$.40}  
&\texttt{1204.66}$\spazio{0.5 } $
 \\ %\hline
  $({\texttt{1}}^{\texttt{1}}{\texttt{p}}_{\texttt{1}})~1^+(1^{+-})~
\phantom{{}^{{}^{\displaystyle{i}}}}\!\!  ${\small{\texttt{b${}_{\texttt{1}}$(1235)}}}
&\texttt{1229.5$\pm$3.2}  
&\texttt{1274.76}
 \\ %\hline
  $({\texttt{1}}^{\texttt{3}}{\texttt{p}}_{\texttt{2}})~1^-(2^{++})~
\phantom{{}^{{}^{\displaystyle{i}}}}\!\!  ${\small{\texttt{a${}_{\texttt{2}}$(1320)}}}
&\texttt{1318.3$\pm$.6}  
&\texttt{1325.40}$\spazio{0.5 } $
 \\ 
%\hline
%
%   2s
%
  $({\texttt{2}}^{\texttt{1}}{\texttt{s}}_{\texttt{0}})~1^-(0^{-+})~
\phantom{{}^{{}^{\displaystyle{i}}}}\!\!  ${\small{\texttt{$\pi$(1300)}}}
&\texttt{1300$\pm$100}  
&\texttt{\phantom{1}1337.36 }
 \\ 
  %\hline
  $({\texttt{2}}^{\texttt{3}}{\texttt{s}}_{\texttt{1}})~1^+(1^{--})~
\phantom{{}^{{}^{\displaystyle{i}}}} \!\! ${\small{\texttt{$\rho$(1450)}}}
&\texttt{1465$\pm$25}  
&\texttt{1497.63}
 \\ 
%\hline
%
%   2d
%
  %\hline
  $({\texttt{1}}^{\texttt{3}}{\texttt{d}}_{\texttt{1}})~1^+(1^{--})~
\phantom{{}^{{}^{\displaystyle{i}}}}\!\!  ${\small{\texttt{$\rho$(1570)}}}
&\texttt{1570${}^{(*)}$}  
&\texttt{1565.42}
 \\ 

%
%   3s
%
  $({\texttt{3}}^{\texttt{1}}{\texttt{s}}_{\texttt{0}})~1^-(0^{-+})~
\phantom{{}^{{}^{\displaystyle{i}}}} \!\! ${\small{\texttt{$\pi$(1800)}}}
&\texttt{1812$\pm$12}  
&\texttt{1882.30}
 \\ 
  %\hline
  $({\texttt{3}}^{\texttt{3}}{\texttt{s}}_{\texttt{1}})~1^+(1^{--})~
\phantom{{}^{{}^{\displaystyle{i}}}}\!\!  ${\small{\texttt{$\rho$(1900)}}}
&\texttt{1900${}^{(*)}$}  
&\texttt{\phantom{2}2016.35}
 $\spazio{0.5}$\\ 
 %\hline
  $({\texttt{2}}^{\texttt{3}}{\texttt{d}}_{\texttt{1}})~1^+(1^{--})~
\phantom{{}^{{}^{\displaystyle{i}}}} \!\! ${\small{\texttt{$\rho$(2150)}}}
&\texttt{2149$\pm$17}  
&\texttt{\phantom{2}2064.36}

$\spazio{0.5}$\\ 
\hline
  \end{tabular}
}}
\label{Table_u-dbar}
\caption{The \texttt{u}\texttt{\={d}} levels in MeV. $\sigma$=$\texttt{1.34}$\,GeV/fm,
$\alpha$=\texttt{0.656}, $m_d$=$\texttt{6.1}\,$MeV, $m_u$=$\texttt{2.94}\,$MeV. 
${}^{(*)}$Meson Summary Table, \cite{pdg}.}
\end{table}

We next consider the $s\bar s$ system, for which there are few accepted experimental states. The much lighter mass of
the $s$ quark
highly enhances the relativistic character of the $s\bar s$ composite system and the fundamental role of the Breit
corrections, giving rise to large hyperfine splittings. Due to these reasons the
string tension $\sigma$ has not been given the same value of the previous systems but has been considered a fitting
parameter, finding a value larger than in $b\bar b$. We report our results in Table III, where we have
also included the unassigned \texttt{f${}_\texttt{1}$(1420)}, $\texttt{X}\texttt{(1750)}$,
$\phi_\texttt{3}\texttt{(1850)}$ and $\phi\texttt{(2170)}$ .  
Although we cannot have a complete phenomenological confidence in the numerical results, still a fair number of
experimental data can be accommodated with a pretty good accuracy. For instance the model could suggest a
$({\texttt{1}}^{\texttt{3}}{\texttt{d}}_{\texttt{1}})$ assignment for $\texttt{X}\texttt{(1750)}$.
We then use the mass of the $s$-quark together with the $b$ and $c$ masses to determine the
levels of the $Bs$ and $Ds$ mesons, reported in Table IV. Even for
different quark masses the agreement with the data is very good.

We finally look at the lightest $u\bar d$ mesons, for which the Breit correction,
as usually calculated, is commonly accepted to be insufficient
to reproduce the data. We have again fitted the data with a constant rcc. The fit
includes also the very light $\rho(770)$, but obviously excludes the $\pi^{\pm}$ for which the use of
a higher $\alpha$ cannot be avoided, due to the steepness of the $\alpha_S$ curve for very low masses. The results are
not very sensitive to the mass ratio $\rho$ that we fix at the physical value 0.35;
the string tension appears to be the same found for $s\bar s$. Finally, the $u$ and $d$  masses are found
close to current algebra masses as opposed to constituent masses (see also \cite{Bra}), normally much higher in
potential models. The exact mass 139.5 MeV of $\pi^{\pm}$ is got with $\alpha=0.99$.

To conclude we give some values of the Breit corrections $\Delta_{q\bar q}$ for different states.
For $1{}^1s_0$ the values in MeV of $(\Delta_{b\bar b},\Delta_{c\bar c},\Delta_{s\bar s},\Delta_{u\bar d})$ respectively
are (92.31, 155.22, 296.81, 600.12). For $1{}^3s_1$, (18.09, 38.80, 94.37, 106.21). For $1{}^3p_2$,
(7.51, 21.10, 55.93, 63.72). 
As expected, the corrections
decrease for increasing values of $j$ and become more and more important for decreasing quark masses. 
We thus believe that the results we have presented show that the covariant formulation based on Dirac equations, 
in addition to being conceptually very simple, is also extremely effective in quarkonium models.

{\textsc{Acknowledgment.} {We would like to thank our colleagues Stefano Catani and Francesco Bigazzi for useful
discussions and interest in our work.}}

\end{document}